\documentclass[a4paper]{spie}  
\usepackage{graphicx}
\usepackage{url}
\usepackage[latin1]{inputenc}
\usepackage[OT1]{fontenc}
\usepackage{multirow}

\title{A publication database for optical long baseline interferometry} 

\author{%
  Fabien Malbet\supit{a\dag}, %
  Guillaume Mella\supit{a\dag}, %
  Peter Lawson\supit{b}, %
  Esther Taillifet\supit{a\dag}, %
  Sylvain Lafrasse\supit{a\dag}, %
\skiplinehalf
\supit{a}Lab.\ d'Astrophysique de Grenoble (LAOG), UMR 5571 Univ.\ J.\ Fourier/CNRS, BP 53, F-38051 Grenoble cedex 9, France;\\
\supit{b}Jet Propulsion Lab., California Institute of Technology, Pasedena, CA 91109, USA;\\
\supit{$\dag$}Jean-Marie Mariotti Center (JMMC), France;
}

\authorinfo{Send correspondence to F.\ Malbet, E-mail: \texttt{Fabien.Malbet@obs.ujf-grenoble.fr}}

\begin{document} 
  \maketitle 

\begin{abstract}
Optical long baseline interferometry is a technique that has generated
almost 850 refereed papers to date. The targets span a large variety
of objects from planetary systems to extragalactic studies and all
branches of stellar physics. We have created a database hosted by the
JMMC and connected to the Optical Long Baseline Interferometry
Newsletter (OLBIN) web site using MySQL and a collection of XML or PHP
scripts in order to store and classify these publications. Each entry
is defined by its ADS bibcode, includes basic ADS informations and
metadata. The metadata are specified by tags sorted in categories:
interferometric facilities, instrumentation, wavelength of operation,
spectral resolution, type of measurement, target type, and paper
category, for example. The whole OLBIN publication list has been
processed and we present how the database is organized and can be
accessed. We use this tool to generate statistical plots of interest
for the community in optical long baseline interferometry. 
\end{abstract}



\section{INTRODUCTION: RATIONALE}
\label{sec:rationale}

Optical interferometry is a technique which requires a high level of
critical subsystems illustrated by the fact that one needs to control
at the nanometer level optical path difference which can reach several
hundred meters, or, to operate several telescopes with some level of
adaptive optics. Furthermore, even for the common professional
astronomer the link between the measurements and the astrophysical
consequences consists in numerous mathematical operations which are
not straightforward to understand. Therefore, despite important
financial and human investments, it seemed for a while that the
astrophysical return was first limited and then restrained to a few
specialized areas even though the gain in spatial resolution is
recognized as a real breakthrough.

The distance between firstly the efforts and the necessary support
from the astronomical community and secondly the results contained in
the peer-reviewed literature both in instrumentation but also for the
astrophysical advances have led the community to get organized and to
publicize its results. This was achieved first by establishing a
common point of reference, the web site OLBIN (Optical Long Baseline
Interferometry Newsletter edited by P. Lawson, see contribution on
this subject in this volume), by forming the IAU commission
\#54\footnote{\texttt{http://olbin.jpl.nasa.gov/iau}} and by tracking
the publication record in the field.

In 2000, the rate of refereed papers published in interferometry of
about 30 papers per year was still manageable by hand but ten years later
this rate has reached about 100 papers/year and it is still growing. The
need to record any new reference in the field is even stronger but it
can no longer be done by hand.  Therefore we have built a database
based on today software capability which allows us to
track the evolution of the field by using new information that add extra
value for the service to the community. 

\section{A bibliographic database directly linked to ADS}

\begin{figure}[p]
  \centering
  \fbox{\includegraphics[height=0.95\textheight]{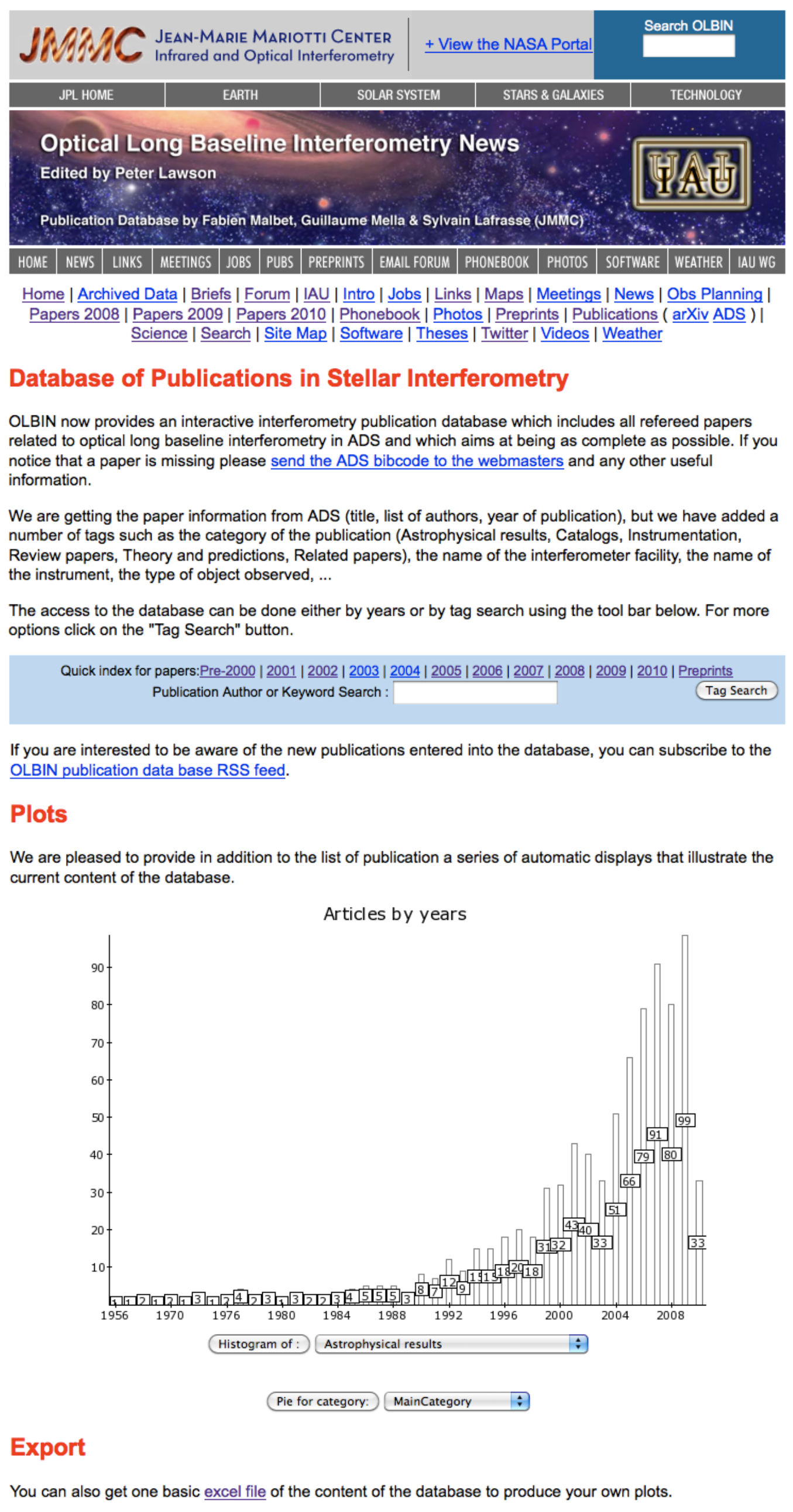}}
  \medskip
  \caption{Homepage of the OLBIN publication database.}
  \label{fig:bibdb-index}
\end{figure}

\begin{figure}[t]
  \centering
  \includegraphics[width=0.95\hsize]{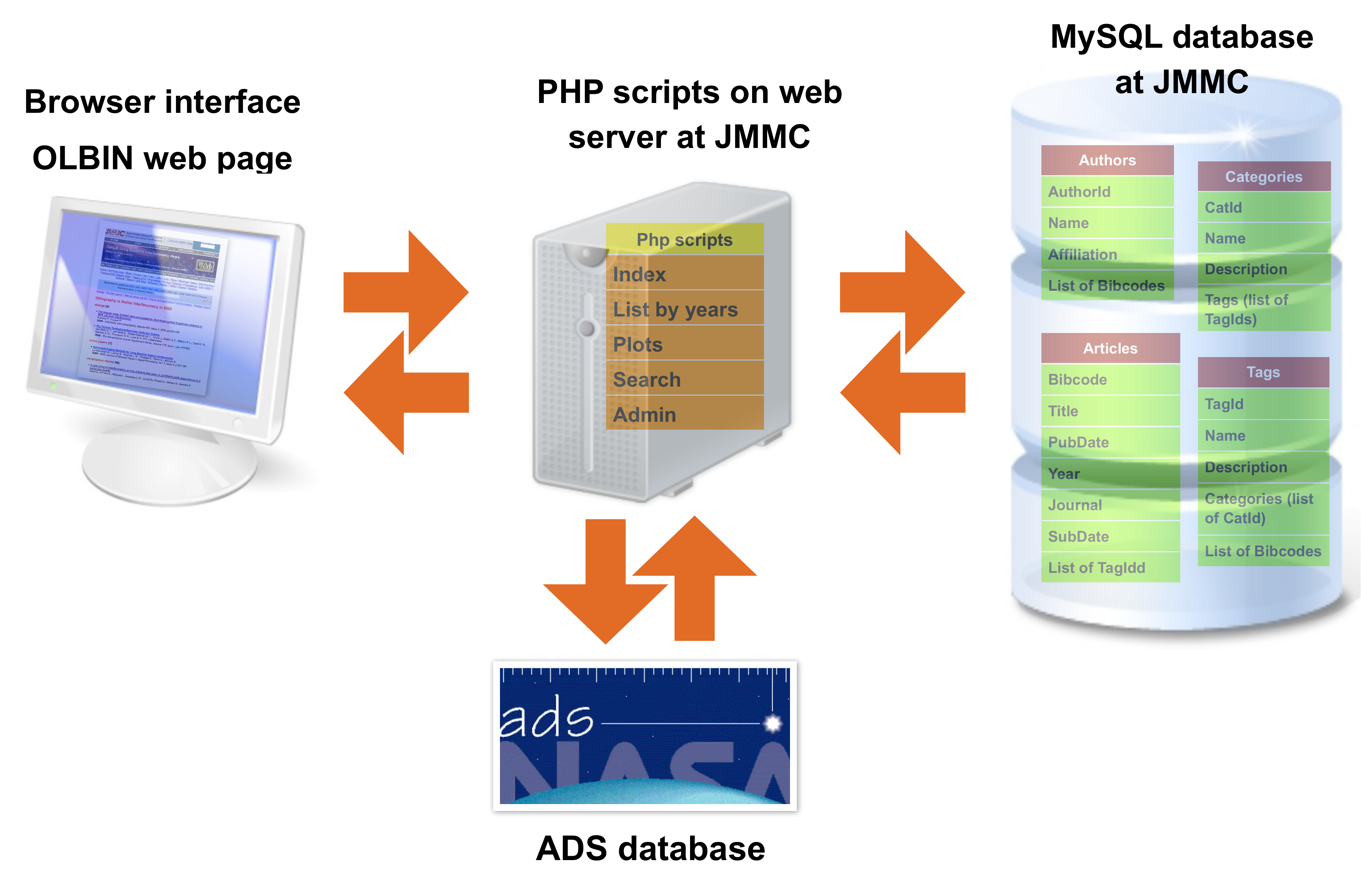}
  \caption{General architecture of the OLBIN publication database}
  \label{fig:diag}
\end{figure}

We decided to restrict ourself to publications which have gone through
the well-recognized peer-reviewed system in order to guaranty the
quality of the papers. As a matter of fact, this type of publication are usually
the ones that agencies but also authors want to refer to.

\subsection{ADS: the primary source of information}

To build this publication database, we want to avoid to enter any
already existing information. In astronomy, we can count on the NASA
\emph{Astrophysics Data System}\cite{2000A&AS..143...41K} (ADS) which
has become the reference in our field covering not only astronomy but
also domains like optics and physics.  In any case, ADS is very open
to inputs from the community through their feedback submission forms
on the Internet. ADS provides many information on the publication,
including the full bibliographic reference, the list of authors, the
year of publications, but also links to the papers on the publisher
sites, to the arXiv preprint database, citation rate and other
useful information.

\begin{figure}[p]
  \centering
  \fbox{\includegraphics[height=0.95\textheight]{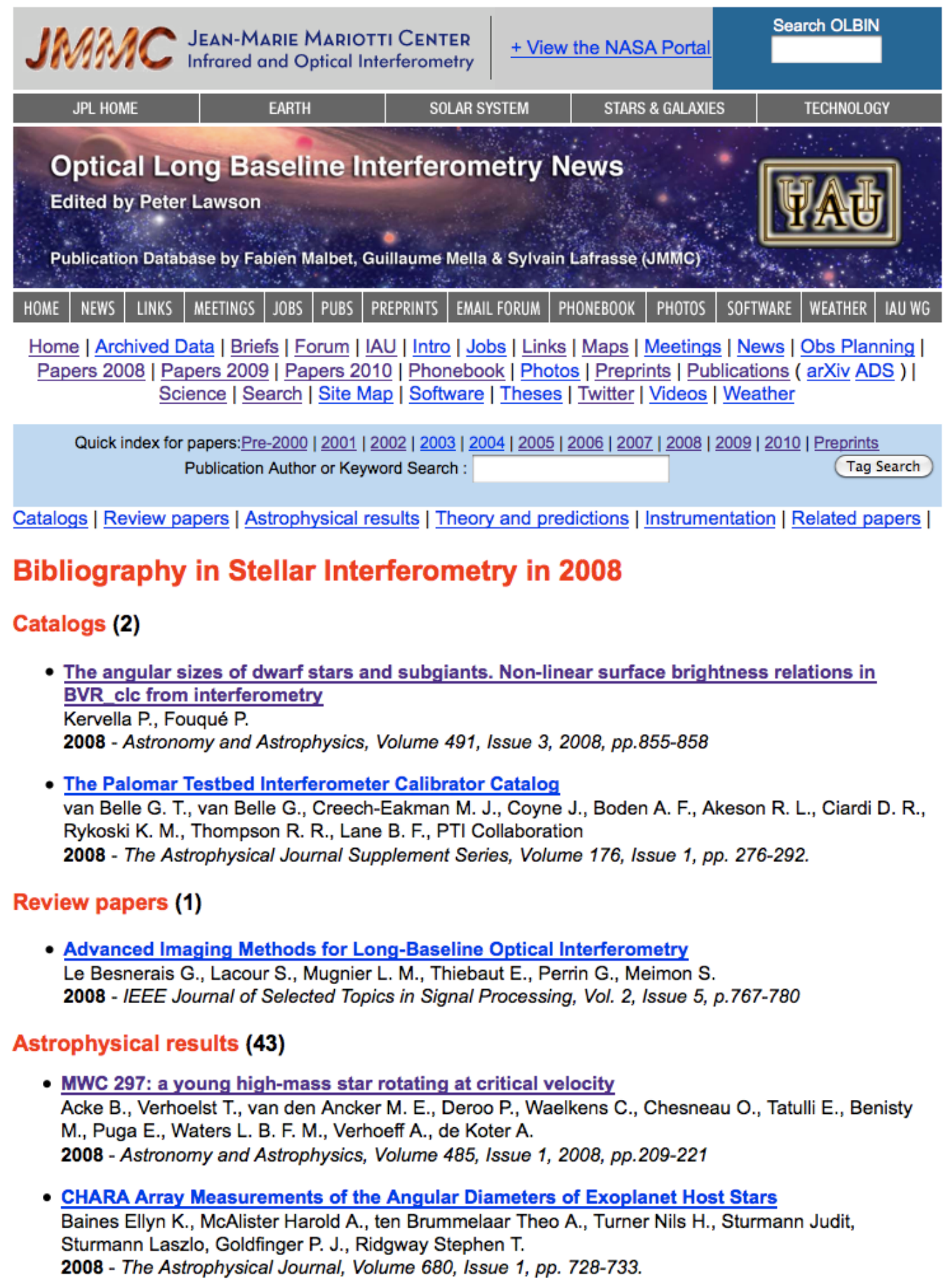}}
  \medskip
  \caption{Lay-out of the OLBIN publication database listing for year 2008.}
  \label{fig:bibdb-2006}
\end{figure}

Our first goal is to provide an up-to-date list of publication ordered
by year of publication and by type of subject (see for example such a
listing in Fig.~\ref{fig:bibdb-2006} for year 2008). Therefore the
output to the user is a web page listing the articles with links to
the ADS abstract page of these papers. The user can access to the
paper they want by just clicking on the link and are redirected to the
ADS pages. There is therefore no need to duplicate the information
stored in ADS.

The central entry in the OLBIN publication database is therefore the
ADS \texttt{bibcode}. The ADS \texttt{bibcode} is a code made of 19
letters which includes the year of publication, the short name of the
publication, the volume, the page and the last name initial of the
first author. The ADS abstract page can be accessed through the HTTP
protocol using the ADS mirrors address completed with the
\texttt{abs/bibcode} link. To enter a new entry in the database, one
needs only to enter the ADS bibcode.

However for classification purposes, the title, the year of
publication, the list of authors, the
affiliations\footnote{Unfortunately, ADS does not provide a way to
  separate the affiliations of the different authors so that this has
  to be handle by our database.} and the bibliographic references
(journal, volume and pages) have to to be stored so that the automatic
generation of the listings can be performed without having to connect
the ADS database.

\subsection{Accessing to the database: dynamical pages}

\begin{figure}[t]
  \centering
  \includegraphics[width=0.95\hsize]{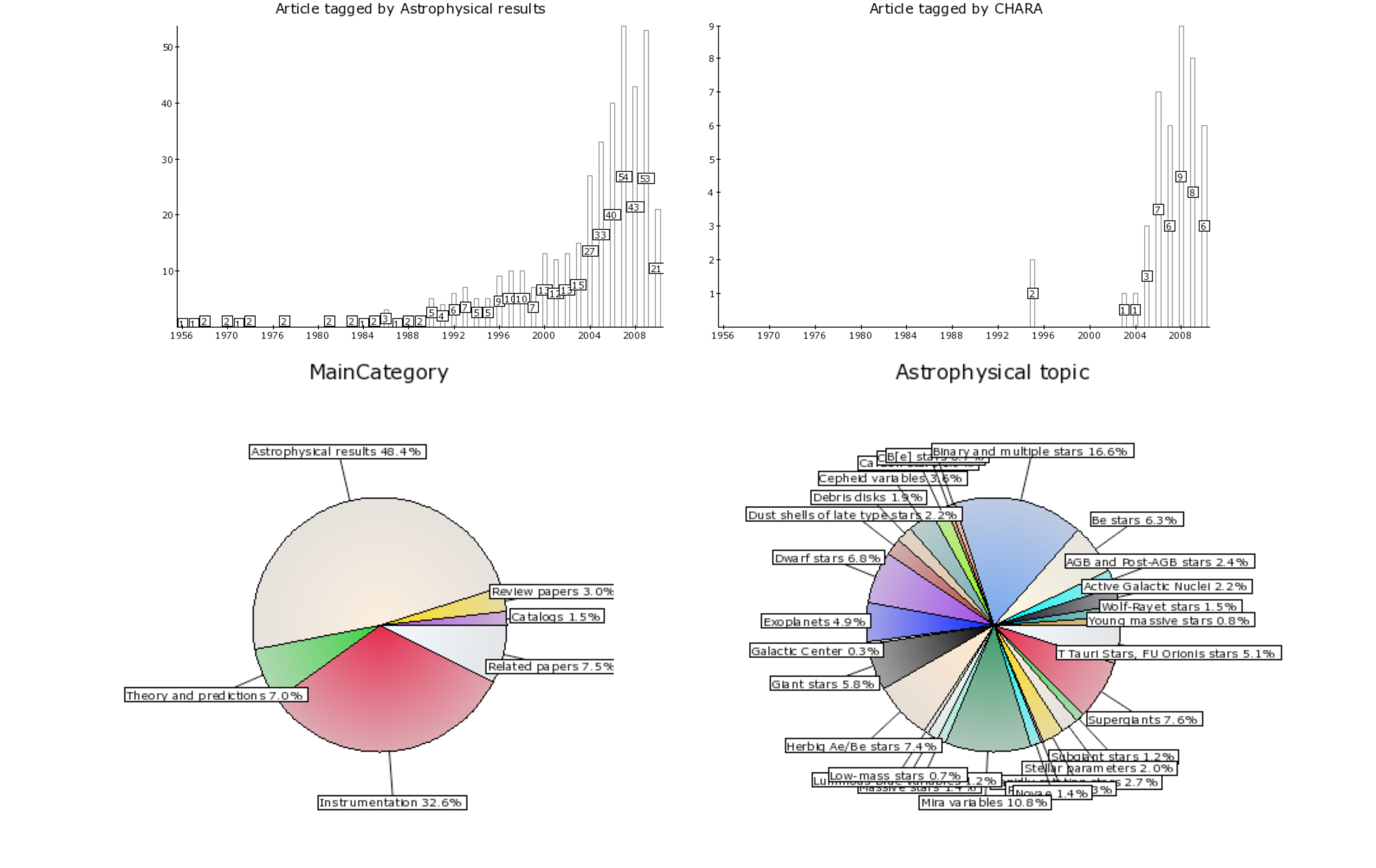}
  \caption{Plots that are automatically from the index OLBIN
    publication homepage. Top: histograms; bottom pie charts.}
  \label{fig:bibdb-plots}
\end{figure}

The main outputs of the database are dynamical pages that can computed
on the fly. This is performed using scripts that can form the HTML
code that is then read by the user's Internet browser.

There are several scripts. Some of them are used only for
administrative reasons. The main scripts are:
\begin{itemize}
\item list of publication for a given year. This is the way the
  information appeared historically on the OLBIN site. We have kept
  this functionality, the only change is that these pages are now
  automatically created upon the user's request. The year list is then
  ordered by type of publication and by alphabetical order of the
  first author.
\item search page which can retrieve all bibliography entries whose
  title, author list contain the searched keys. We can also specify
  in this list the tags to be searched. A combination of tags
  restricts the list to the entries which contains these tags.
\item plots automatically generated with tags. Two types of plots can
  be produced: histogram of number of papers published per year for a
  given tag, or, pie chart based on the category of tags (see Fig.~\ref{fig:bibdb-plots}).
\item access to the database content under the form of a
  comma-separated-values (CSV) file which can be analyzed with usual
  spreadsheet software.
\item administrative pages solely accessible by the OLBIN editors
  which allow them to add new entries by providing the bibcodes and
  checking the associated tags, to manage the list of tags and categories of
  tags, to search the database for entries and manage the checked
  tags, to delete an entry. 
\end{itemize}

\begin{figure}[t]
  \centering
  \includegraphics[width=0.7\hsize]{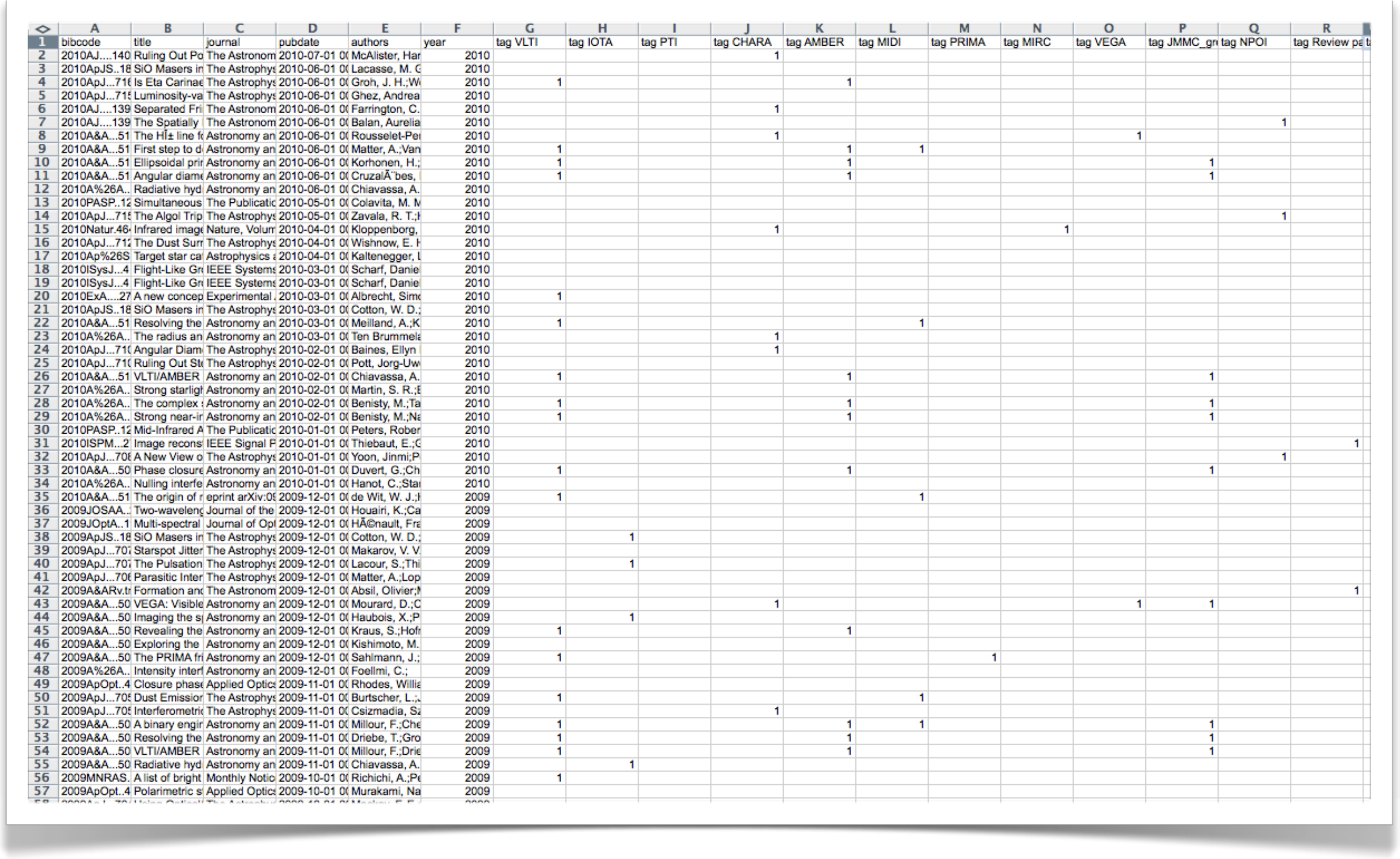}
  \caption{Comma-separated-values (CSV) file that is automatically from the index OLBIN
    publication homepage which can be processed by any spreadsheet software.}
  \label{fig:bibdb-csv}
\end{figure}

One can check the tags associated with an entry by putting the mouse
pointer over the entry link in all dynamical pages generated by the
database.  

Since we are using ADS, the first thing to do when entering a paper is to
check that it is in ADS and, if not, we have to request it to be
registered. Similarly, if mistakes are found, then they should be
corrected in ADS since it is the most used publication database in
astronomy.  

\subsection{Metadata: publication tags} 

In order to add extra-values to the database and to provide other way
of classifying the resulting lists, we added the notion of \emph{tags}
to each entry to better define the different entries. Any paper can be
labeled by any number of tags. In order to sort out the database, we
also created \emph{categories} of tags which are lists of tags of same
nature: type of papers, facilities, instruments, astrophysical topics,
technique,... 

At the time being, there are about 70 tags which label the papers in the
OLBIN database. We have classified them into 7 categories:
\begin{itemize}
\item \textbf{Type of papers}: Astrophysical results, Catalogs, Instrumentation, 
Related papers, Review papers, Theory and predictions;
\item \textbf{Facility}: CHARA, COAST, GI2T, I2T, IOTA, IRMA, ISI,
  Keck, LBTI, Mark III, Narrabri Stellar Intensity Interferometer
  (NSII), NPOI, PTI, SIM, SUSI, VLTI;
\item \textbf{Instrument}: AMBER, CHARA Classic, FLUOR, IONIC, MIDI, MIRC, PRIMA,
VEGA, VINCI; 
\item \textbf{Astrophysical topic}: AGB and Post-AGB stars, Active
  Galactic Nuclei, B[e] stars, Be stars, Binary and multiple stars,
  Calibrators, Carbon stars, Cepheid variables, Debris disks, Dust
  shells of late type stars, Dwarf stars, Exoplanets, Galactic Center,
  Giant stars, Herbig Ae/Be stars, Low-mass stars, Luminous Blue
  Variables, Massive stars, Mira variables, Novae, R CrB stars,
  Rapidly rotating stars, Stellar parameters, Subgiant stars,
  Supergiants, T Tauri Stars, FU Orionis stars, Wolf-Rayet stars,
  Young massive stars;
\item \textbf{Wavelength of operation}: visible, near-infrared (NIR),
  mid-infrared (MIR);
\item \textbf{Spectral resolution mode}: broad-band,
  narrow-band , low resolution ($\mathcal{R}\leq100$), medium resolution
  ($100\leq \mathcal{R} \leq 3000$), high resolution ($3000\leq\mathcal{R}$); 
\item \textbf{Technique}: Astrometry, Closure phases, Differential phase, Fringe
tracking, Images, Intensity interferometry, Nulling, Phase reference,
Squared visibilities,...
\end{itemize}

Adding tags must be easy as well as reorganizing the categories of
tags. However, one should keep in mind that when adding tags, the
whole database should be scanned again in order to make sure that no
entries have been forgotten. This is the main weakness of this architecture.

\subsection{Technical solution}

Figure~\ref{fig:diag} represents the technical architecture which has
be chosen to build the OPLBIN publication database. This database is
based on \texttt{MySQL} and it is located in the
JMMC\footnote{Jean-Marie Mariotti Center, \texttt{http://www.jmmc.fr}}
server in Grenoble. This database contains different tables with the
following information attached:
\begin{itemize}
\item \textbf{Articles}: \texttt{Bibcode}, \texttt{Title}, \texttt{PubDate}
  corresponding to the publication date, \texttt{Year}, \texttt{Journal},
  \texttt{SubDate} corresponding to the date of submission in the
  database, list of \texttt{TagId} which identify to the tags.
\item \textbf{Authors}: \texttt{AuthorId}, \texttt{Name} of the
  author, \texttt{Affiliation} of the author, list of attached
  \texttt{Bibcode}.
\item \textbf{Tags}: \texttt{TagId} which uniquely identify the tag,
  \texttt{Name} of the tag, \texttt{Description} of the tag (not
  always filled in), list of attached \texttt{CatId}, list of attached \texttt{Bibcode}.
\item \textbf{Categories}: \texttt{CatId}  which uniquely identifies the
  category, \texttt{Name} of the category, \texttt{Description} of the
  category, list of attached \texttt{TagId} which are gathered in this category.
\end{itemize}

The \texttt{MySQL} database can be accessed through XML or PHP scripts
which construct dynamically the pages that can been read using a
browser. Input data can also be introduced through forms in order to store
new information mainly in administrative mode. XML scripts are used to
interact with ADS when a new entry is given.

\section{Content of the OLBIN publication database }
\label{sec:cont-olbin-publ}

\begin{figure}[t]
  \centering
  \includegraphics[width=0.8\hsize]{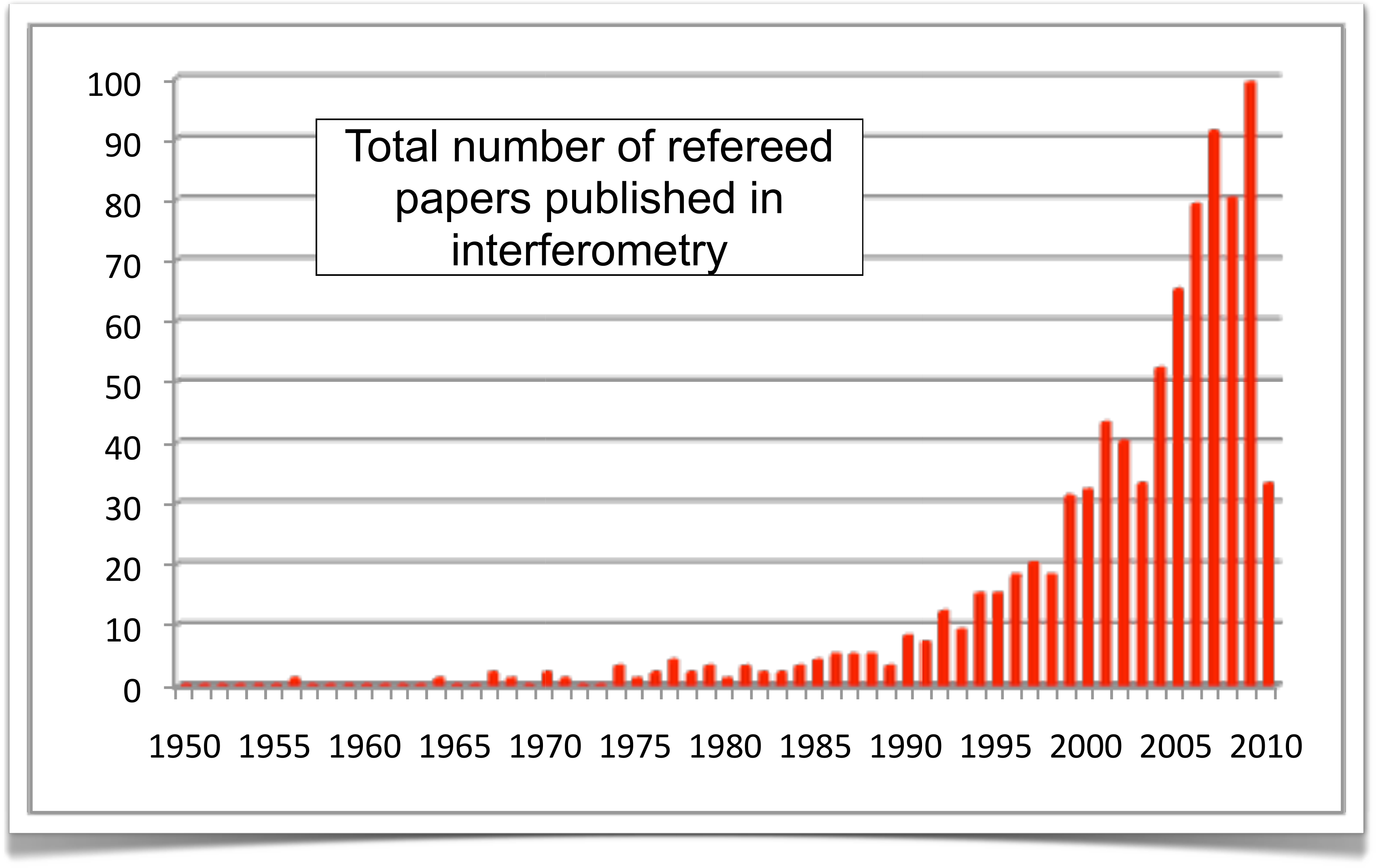}
  \caption{Total number of refereed publications published since 1950.}
  \label{fig:bibdb-total-years}
\end{figure}
 
In this Section, we would like to give examples of applications concerning the
publication in interferometry. They originate directly from the data
base and the use of the CSV export file.

\subsection{Growth of interferometry results}

\begin{figure}[t]
  \centering
  \includegraphics[width=0.8\hsize]{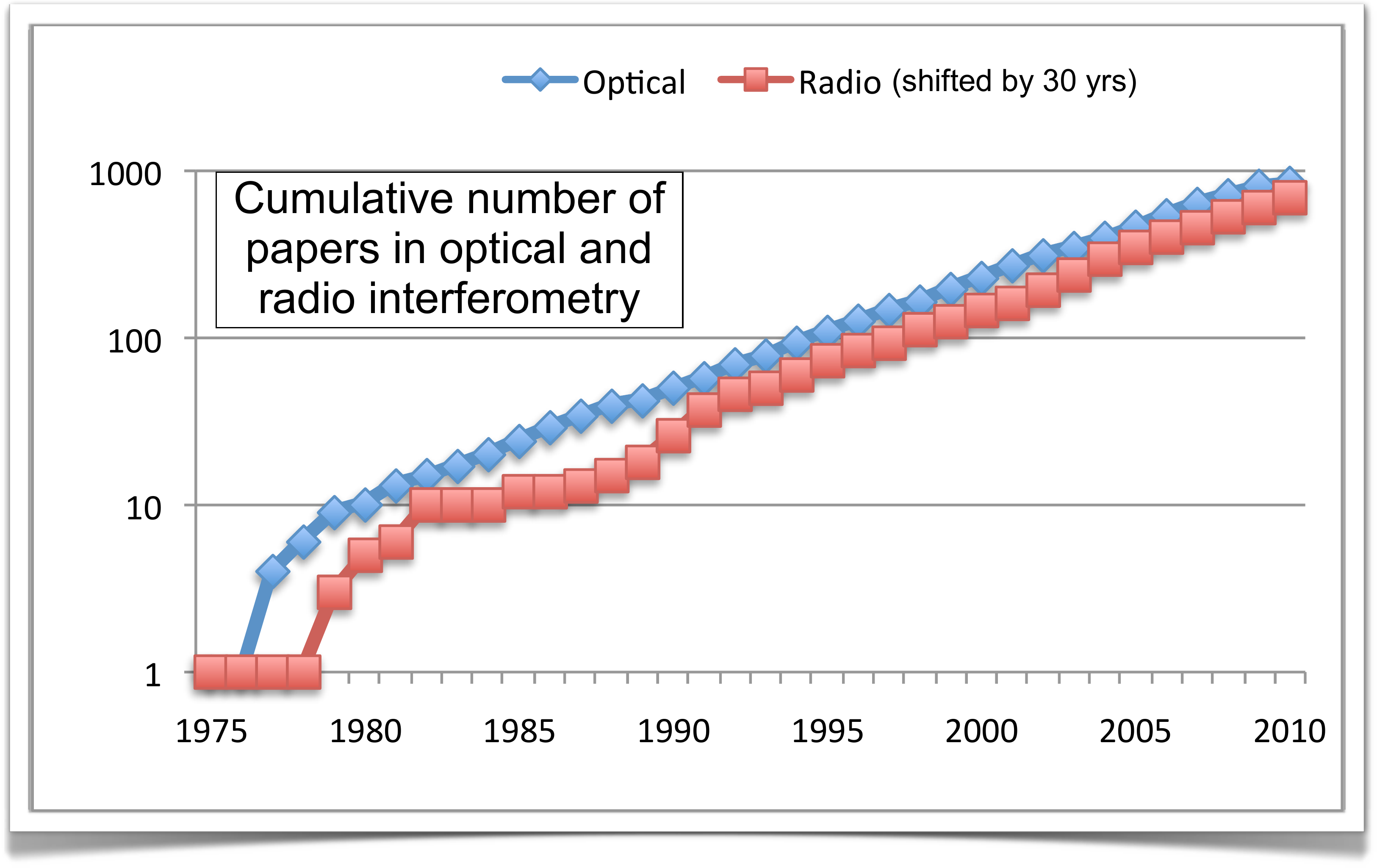}
  \caption{Evolution of the cumulative number of refereed publications
    published since 1950 between the optical (blue losanges) and the
    radio (red squares) domains.}
  \label{fig:bibdb-cumulative-vs-radio}
\end{figure}

At the date of presentation of this work, there is a total of more
than 850 refereed publications in the database. There is also about 70
tags and 7 categories. The evolution of the number of publications is
displayed in Fig.~\ref{fig:bibdb-total-years}. The growth is
exponential with a rate of publication less than 5 papers per year in
the 1980's that reached 30 papers per year in 2000 and almost 100
papers per year in 2010.

An interesting way to analyze this growth is to compare it to the
growth of papers produced by radio-interferometers at the beginning of
the domain. Figure~\ref{fig:bibdb-cumulative-vs-radio} shows the
cumulative number of papers in optical and infrared interferometry
(blue losanges) compared to radio-interferometry (red squares). If we shift the radio
curve by 30 years then it becomes obvious that optical interferometry
follows the same type of growth of radio-interferometry.

\begin{figure}[t]
  \centering
  \includegraphics[width=0.55\hsize]{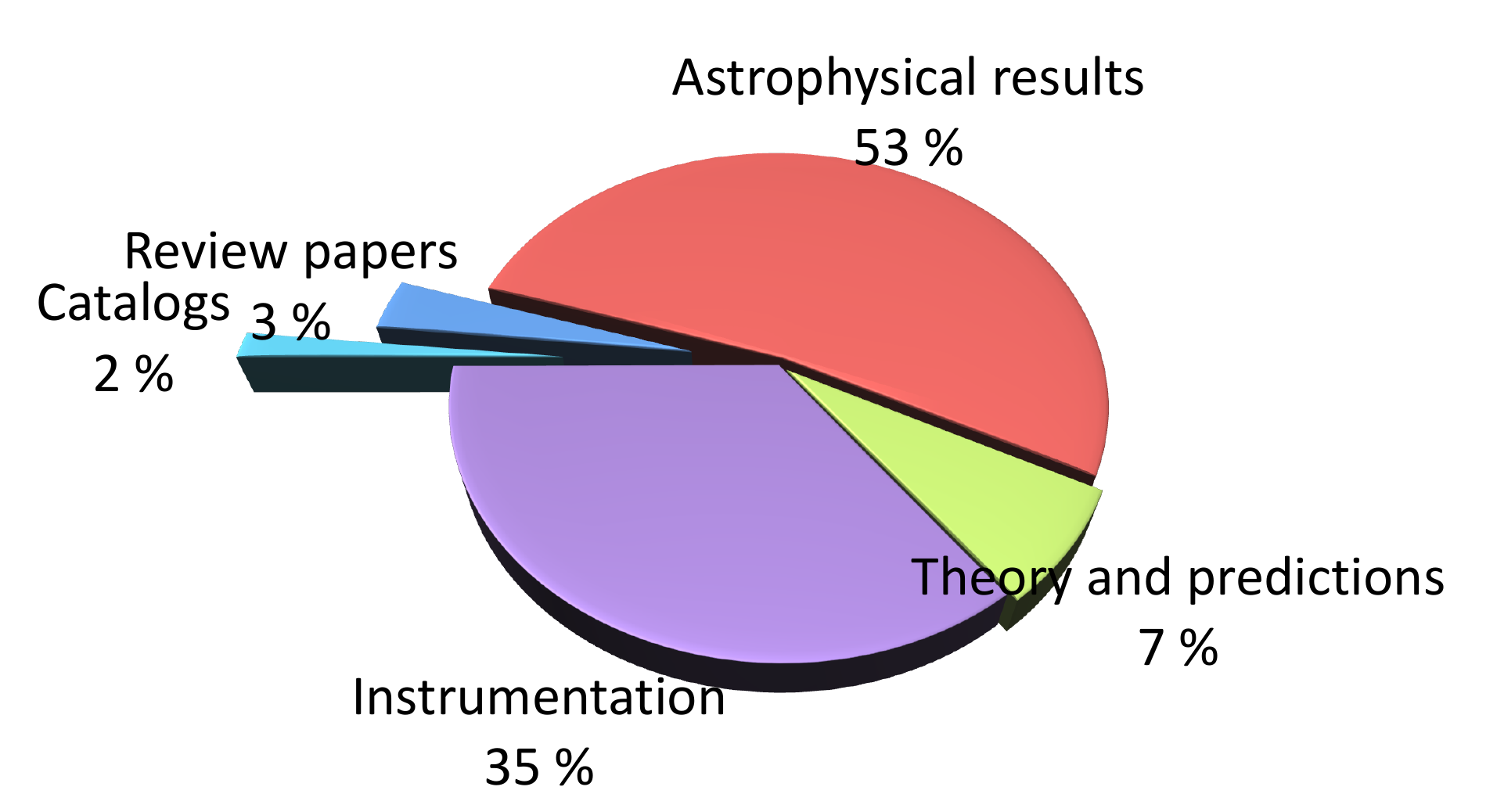}
  \caption{Repartition of the different types of refereed articles. }
  \label{fig:bibdb-paper-categories}
\end{figure}

The type of papers is presented in Fig.~\ref{fig:bibdb-paper-categories}. More
than half the papers correspond to actual astrophysical results, 35\%
of them are articles on instrumentation and concept and in the
remaining 10\% we have theory and predictions, catalogs and reviews.

\subsection{Astrophysical topics}

\begin{table}[t]
  \centering
  \caption{Tags of the main category \emph{astrophysical topics} sorted
    in intermediate astrophysical sub-categories. } 
  \label{tab:categ}
  \bigskip
  \begin{tabular}{|l|l|l|l|}
    \hline
    \textbf{Stellar parameters}	&\multicolumn{2}{c|}{\bf Evolved
      stars} 
    &\textbf{Hot active stars}\\
    \hline
    Supergiants 	&R CrB stars 	&Mira variables  &Massive stars 	\\

    Giant stars 	&Wolf-Rayet stars 	&Cepheid variables
    &Luminous Blue Variables \\

    Subgiant stars 	&Novae 	&Carbon stars 	&B[e] stars 	\\

    Dwarf stars 	&Dust shells of late	&AGB
    and Post-AGB 	&Be stars	\\ 

    Rapidly rotating stars &-type stars&&		\\
    \hline
    \multicolumn{4}{c}{}    \\
    \hline
    \textbf{Young stellar objects}	 &\textbf{Low-mass objects}	&\textbf{Galaxies}	&\textbf{Multiple stars}\\
    \hline
    T Tauri Stars, FU Orionis &Low-mass stars &Galactic Center 
    &Binary and multiple systems \\ 

    Herbig Ae/Be stars &Exoplanets         &Active Galactic Nuclei  &	\\ 

    Young massive stars	&Debris disks 	&&	\\ 
    \hline
 \end{tabular}
\end{table}

\begin{figure}[t]
  \centering
  \includegraphics[width=0.9\hsize]{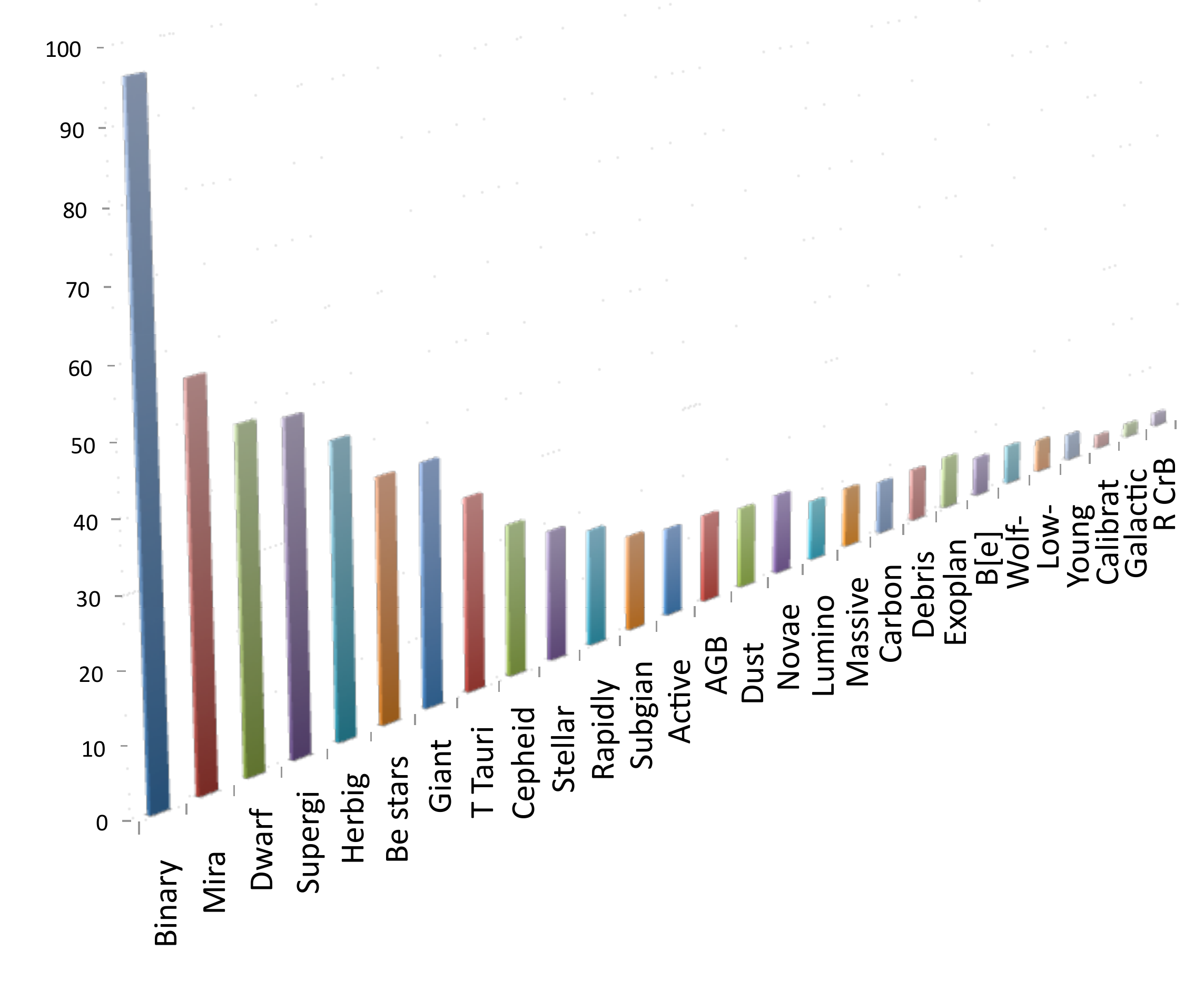}
  \caption{Repartition of the number of publications per topics.}
  \label{fig:bibdb-topics}
\end{figure}
\begin{figure}[t]
  \centering
  \includegraphics[width=0.7\hsize]{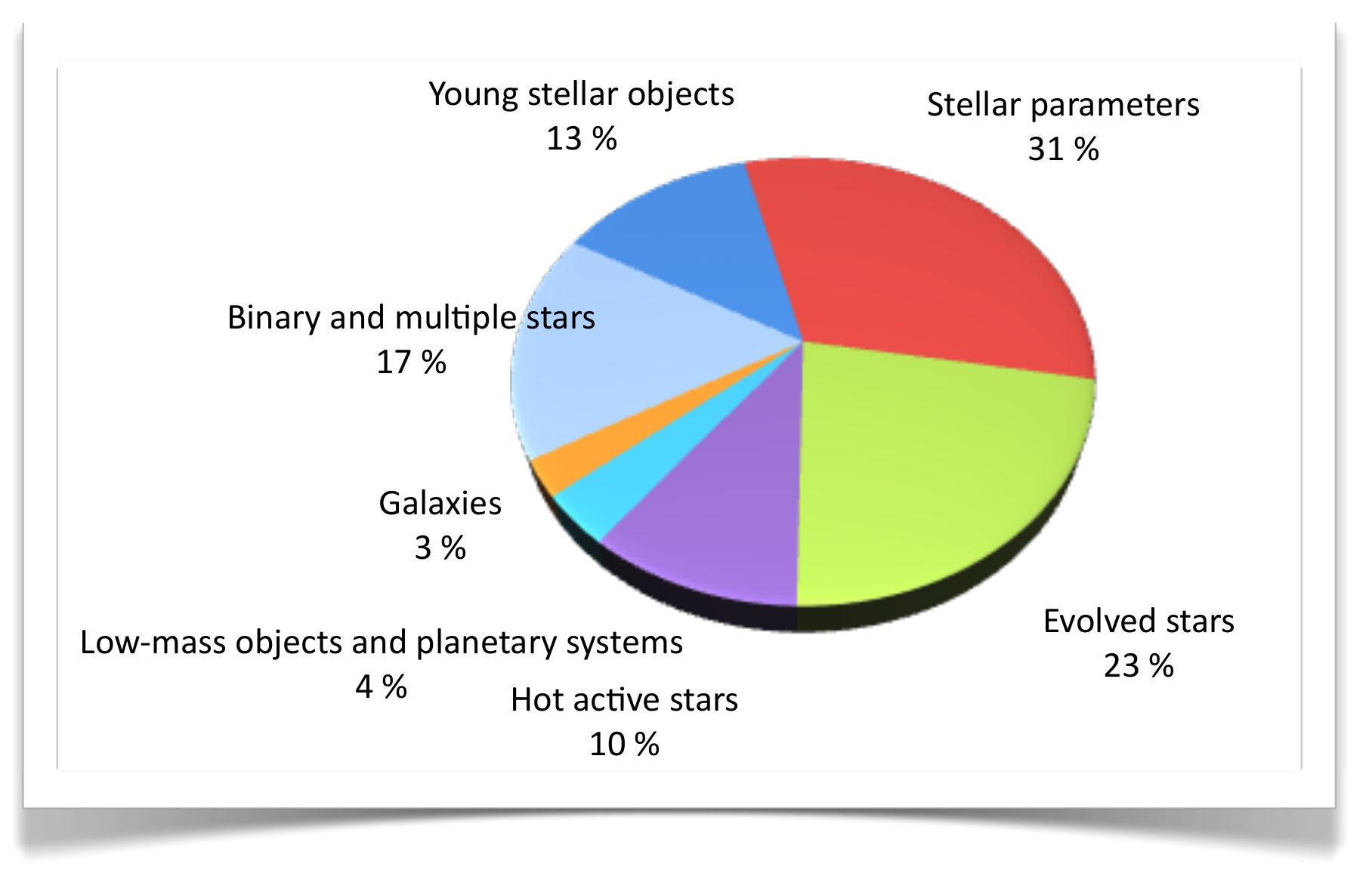}
  \caption{Pie chart of the main science categories of the OLBIN
    publication database.}
  \label{fig:bibdb-topics-main}
\end{figure}

Figure \ref{fig:bibdb-topics} shows the repartition of papers
published per topics in decreasing order. The number of topics are too
detailed to distinguish a particular trend. We have therefore defined
general astrophysical categories which gather several topics (see
Table~\ref{tab:categ}). Figure \ref{fig:bibdb-topics-main} summarizes
the share of the different astrophysical main categories. The largest
category concerns the stellar parameters (31\%) which is not
surprising since it is the first science that has been achieved by
optical interferometry. The second largest category consists of
evolved stars (23\%) which have benefited from the high brightness of
some of them. Multiple systems (17\%) are playing an important role as
well, whereas young stellar objects constitute the fourth category
(13\%) which has been growing a lot during this last decade. The
remaining stellar categories are hot active stars and low-mass objects
and planetary systems. The last objects are more difficult to observe. Finally
extragalactic observations already amount to 3\% of the total number
of astrophysical results.

\subsection{Facilities}

\begin{figure}[t]
  \centering
  \includegraphics[width=0.75\hsize]{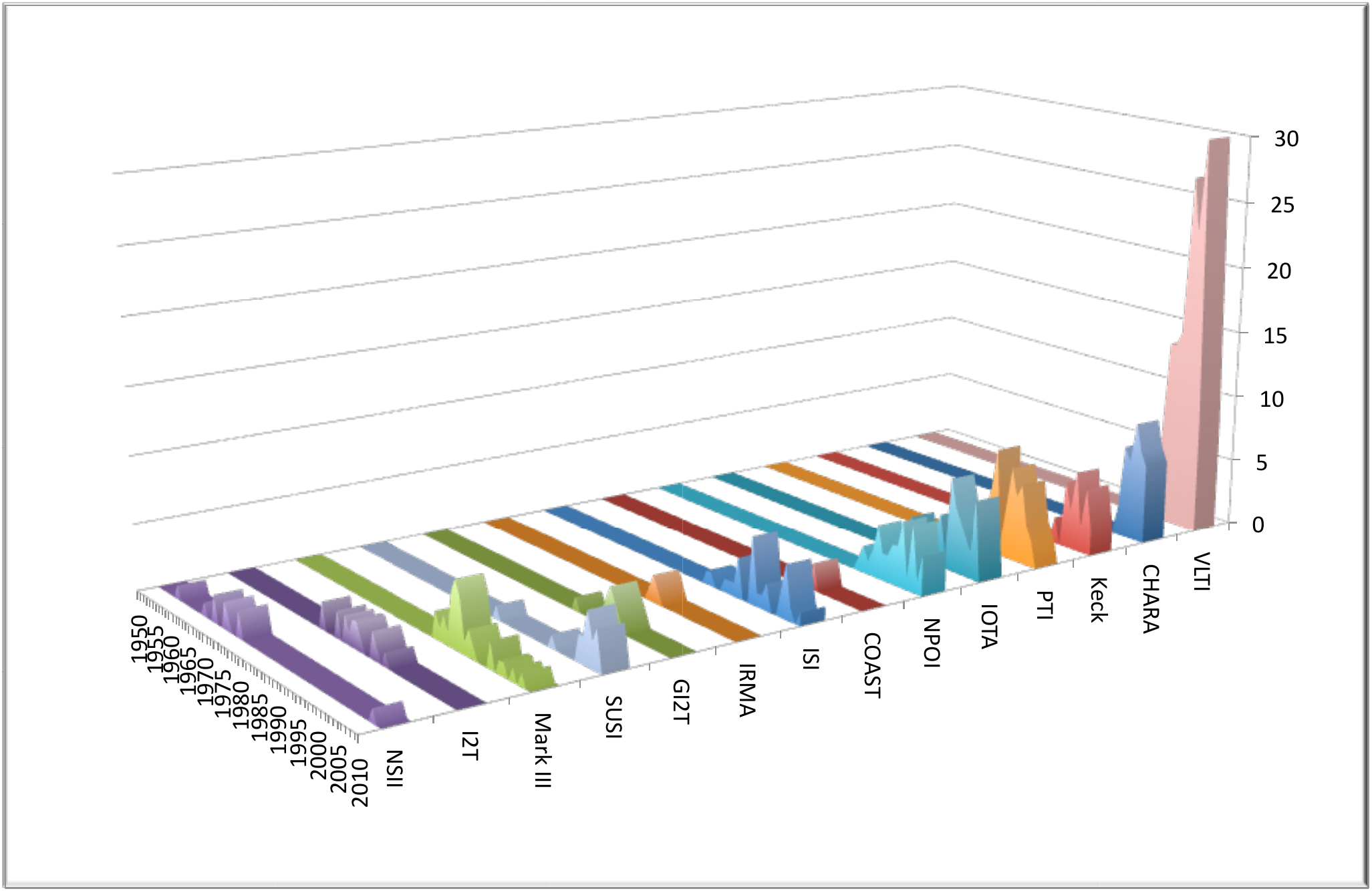}
  
  \caption{Evolution of the number of publication which have been
    tagged to a past or operating facility.}
  \label{fig:bibdb-facilitiess}
\end{figure}

Figure \ref{fig:bibdb-facilitiess} shows the evolution of
astrophysical results published per facility. One can notice the
historical facilities (NSII, I2T, Mark III, IRMA, COAST, GI2T), the
more modern installations (ISI, SUSI, NPOI, IOTA, PTI) to the latest
large facilities (Keck, CHARA and VLTI). The fact that the VLTI is
offered to a large community boosts its number of publication.

\begin{figure}[t]
  \centering
  \includegraphics[width=\hsize]{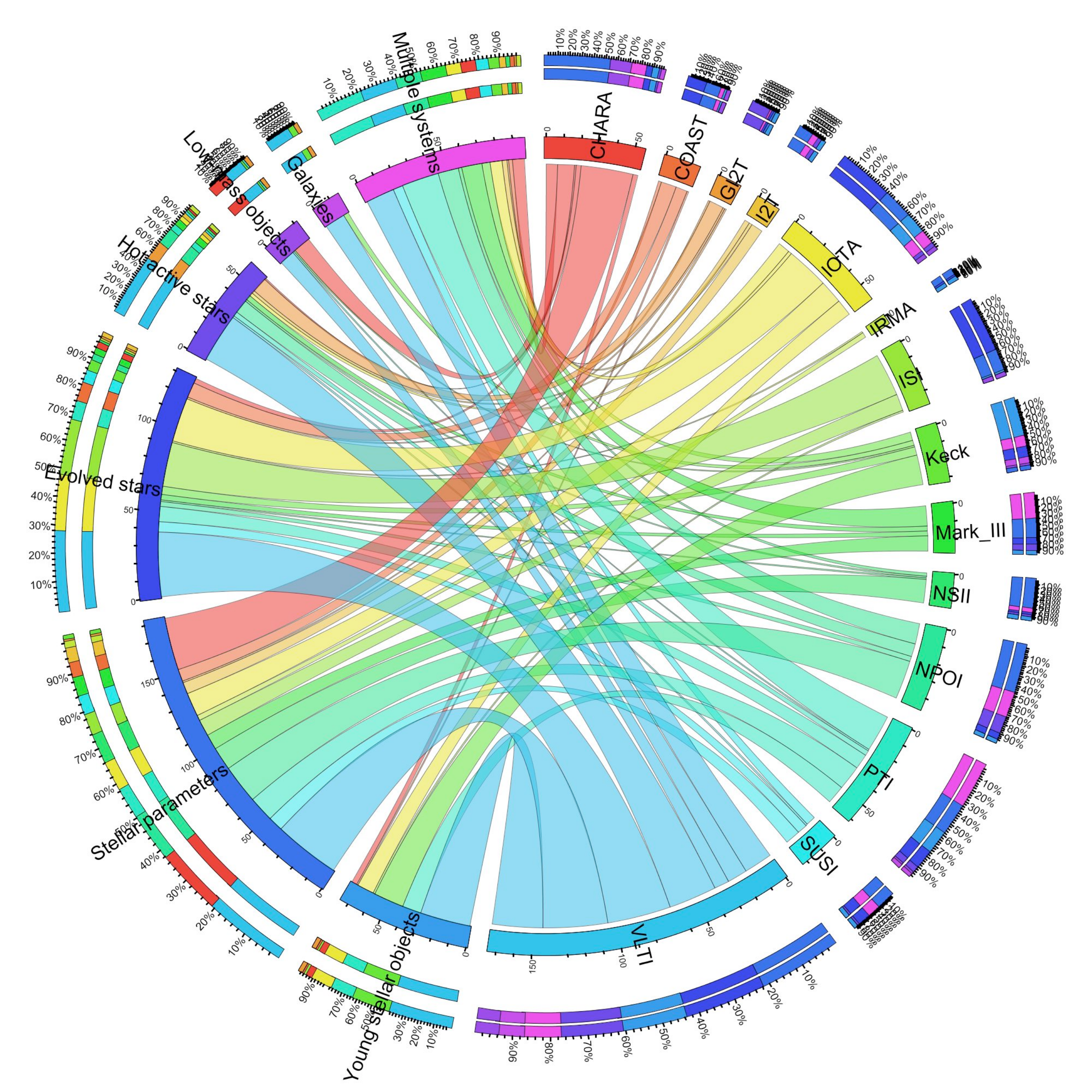}
  \caption{Table of the astrophysical topic vs the facility
    represented with the Circos Table viewer (see
    \texttt{http://mkweb.bcgsc.ca/circos}).}
  \label{fig:bibdb-circos}
\end{figure}

The diagram displayed in Fig.~\ref{fig:bibdb-circos} is particularly
interesting. It displays the correspondance between two categories:
the astrophysical topics (left) and the facilities (right). For
example, VLTI has a rather broad range of topics compared to ISI. On
the other side, low-mass objects have been tackled mainly by CHARA and
the VLTI.

\section{Conclusion and perspectives}
\label{sec:perspectives}

We have described in this paper the implementation of a publication
database to list all refereed publication in optical long baseline
interferometry. We have included some statistical plots which
illustrate the possible applications of this database. We need the
feedback of the users to correct the citations, the tags and be aware
of all papers.

This tool could be the reference for the different groups in order to
list their publications. We may develop then specific pages
(instruments, interferometers, science,..). These pages might also be
used by our agencies to evaluate the outcome of interferometry. We
could also contemplate to get the citations rate from ADS, but since
it changes everyday basically, it would require to update the database
on a daily basis. Another important perspective is to create a direct
link between the publications and the published data.

Finally do not forget to use this database in your work:
\begin{center}
  \fbox{%
    \begin{minipage}[c]{0.8\linewidth}
    \begin{center}
      \textbf{\tt http://olbin.jpl.nasa.gov} in the \emph{Publications} menu\\
      or\\
      \textbf{\tt http://www.jmmc.fr/bibdb/}
    \end{center}  
    \end{minipage}
  }%
\end{center}
\acknowledgements

We would like to thanks all our colleagues who have discussed with us
the opportunity of such a tool. We are especially grateful to S.~Ridgway who
compiled the number of papers published in radio-interferometry.

\bibliography{2010-olbin-bibdb-spie}   

\begin{thebibliography}{1}

\bibitem{2000A&AS..143...41K}
{Kurtz}, M.~J., {Eichhorn}, G., {Accomazzi}, A., {Grant}, C.~S., {Murray},
  S.~S., and {Watson}, J.~M., ``{The NASA Astrophysics Data System:
  Overview},'' {\em \aaps}~{\bf 143},  41--59 (Apr. 2000).

\end{thebibliography}
\bibliographystyle{spiebib}   

\end{document}